# Electro-optic comb-driven massive data transmission on monolithically integrated lithium niobate photonic circuits

Yikun Chen[1,2,5], Chuang Xu[3,5], Ke Zhang[1,2,4], Xiong Wu[3], Kangping Zhong[3], Xiangzhi Xie[1,2], Hanke Feng[1,2], Yifan Wu[1,2], Chao Lu[3], Alan Pak Tao Lau[3,*] and Cheng Wang[1,2,*]

[1] Department of Electrical Engineering, City University of Hong Kong, Kowloon, Hong Kong SAR, China
[2] State Key Laboratory of Terahertz and Millimeter Waves, City University of Hong Kong, Kowloon, Hong Kong SAR, China
[3] Photonics Research Institute, Department of Electrical and Electronic Engineering, The Hong Kong Polytechnic University, Hong Kong SAR, China
[4] RhinoptiX Technology Ltd., Hangzhou, China
[5] These authors contributed equally to this work
[*] Corresponding authors: eeaptlau@polyu.edu.hk; cwang257@cityu.edu.hk

**Abstract**

Driven by large AI models, the demand for datacenter interconnects is outstripping traditional scaling laws. This necessitates a transition toward dense multiplexing techniques, such as wavelength-division multiplexing (WDM), within the datacenter environment, effectively migrating long-haul spectral efficiency to short-reach, high-bandwidth applications. Integrated optical frequency combs could provide compact and cost-effective light sources for future datacenter WDM interconnects. However, current demonstrations based on Kerr combs or mode-locked lasers face critical limitations in power efficiency, spectral flatness or phase noise. Their fixed, device-dependent channel spacings remain incompatible with standard WDM grids, while their integration with high-speed modulators presents a significant bottleneck. Here, we address these challenges by demonstrating massive data transmission using flat-top electro-optic combs on thin-film lithium niobate (TFLN)—a platform distinguished by its flat spectral profile, versatile operating wavelengths and comb spacing, while enabling seamless integration with (de-)multiplexers and high-speed modulators. Using 46 wavelength channels generated by two lasers and a 50-GHz on-chip electro-optic comb, we achieve 16.56 Tb/s coherent data transmission over a 98.84-km field-deployed link connecting two universities and two datacenters in Hong Kong, validating its practical utility for inter-datacenter links. To showcase the WDM grid flexibility of our scheme, we further interleave the comb into 100-GHz spacing via an on-chip de-multiplexer, demonstrating 10-channel, 80-GBaud coherent data transmission at 6.4 Tb/s. Finally, we present a proof-of-concept on-chip WDM transmitter by monolithically integrating the electro-optic comb with a high-speed modulator, delivering 19-channel, 45-GBaud intensity-modulation direct-detection (IM/DD) data transmission at a total bit rate of 1.71 Tb/s. Our results establish TFLN-based electro-optic comb as a highly flexible, compact, and scalable WDM solution for both coherent and IM/DD scenarios, meeting the escalating demands of datacenter interconnects.

## Introduction

The recent surge in large-scale artificial intelligence (AI) models is driving a dramatic increase in traffic across data center interconnections (DCI), with a compound annual growth rate (CAGR) exceeding 30%[1]. To keep pace with this escalating demand, optical transceivers must double their capacity every 2-3 years. To date, this has been achieved through a combination of scaling methods, including increasing symbol rates (e.g. 224-GBaud) based on broader-band components, adopting higher-order modulation format [e.g. transitioning from non-return-to-zero (NRZ) to 4-level pulse amplitude modulation (PAM-4)], and running multiple fibers in parallel [e.g. four lanes in quad small form factor pluggable (QSFP) module][2].

These scaling strategies, however, are becoming increasingly difficult to sustain due to fundamental physical limitations. Further increasing symbol rates beyond 224-GBaud encounters hard bandwidth constraints of core components, such as electro-optic (EO) modulators, photodetectors (PD), digital signal processors (DSP), and radio-frequency (RF) cables. Adopting even higher-order modulation formats provide diminishing returns, as data capacity only increases logarithmically with the number of modulation levels, while at the cost of significantly higher signal-to-noise ratios (SNR) demand. Deploying more fibers is constrained by space limitations in data centers and incurs higher costs for installation and maintenance. These challenges highlight the urgent need for an alternative scaling path to further expand DCI capacity based on compact, low-cost, and scalable technologies[3].

Wavelength-division multiplexing (WDM) is a compelling solution for future DCI applications, leveraging wavelength as an additional encoding degree of freedom[3,4]. However, conventional WDM systems typically rely on discrete laser arrays to power multiple wavelength channels, leading to large footprints, high costs, increased system complexity, and excessive power consumption. As a result, while WDM systems have been widely adopted in long-haul telecommunication networks, existing technologies fall short of meeting the stringent cost, footprint, and power consumption requirements of DCI applications.

In recent years, integrated optical frequency combs such as Kerr solitons[5–18] and semiconductor mode-locked lasers[19–23] have emerged as promising alternative light sources for WDM systems, thanks to their compact size, low cost, and broad spectral coverage. For instance, ultrahigh-capacity data transmission at 55 Tb/s has been achieved based on a 179-channel Kerr comb[13]. However, Kerr combs are plagued by low pump-to-comb conversion efficiency and a lack of spectral flatness, necessitating bulky waveshapers[9,12,16] for spectral flattening in WDM experiments. While Mode-locked lasers offer higher power per line and better spectral flatness[19–23], they typically exhibit narrow bandwidth and inferior noise performance. Importantly, the comb spacings in both approaches are rigid and device-dependent, making alignment with commercial WDM grids challenging in practical deployments[6–9,12–18,20–23] (see Table 1 in Supplementary Information for details), where fabrication variations and environmental perturbations are present. Moreover, these photonic integration platforms are incompatible with other key components required for a full WDM transmission system, such as (de-)multiplexers and/or high-speed EO modulators. The limited platform scalability significantly hinders their potential for realizing full WDM transmitters in compact form factors.

Here, we propose addressing these issues by using integrated EO combs on the thin-film lithium niobate (TFLN) platform. EO combs based on lithium niobate (LN) modulators are well known for their exceptional flexibility, stability, and the ability to generate flat-top spectra via cascaded amplitude and phase modulation[24]. Recent transition from bulk to TFLN platform further endows this seminal material with sub-wavelength optical confinement and enhanced EO interaction, enabling dramatically lowered half-wave voltages[25] and broader EO bandwidths[26] for efficient comb generation[27–32]. Moreover, the EO comb sources can be monolithically integrated with a variety of components, such as (de-)multiplexers[33–35] and high-speed modulators[25,26,36–38], leading to high-performance functional systems on TFLN chips[32,39–43]. Despite these advances and the integration prospects for full WDM transmitter, current integrated EO comb-based data transmission demonstrations[44,45] remain constrained to limited spectral coverage ($\leq$ 0.35 THz without further fiber-based spectral broadening), insufficient comb spacing ($\leq$ 25 GHz), and the requirement of bulky elements like waveshapers and discrete modulators.

We fulfill these unmet demands through monolithically integrated TFLN photonic circuits that incorporate EO combs, wavelength de-multiplexers, and high-speed EO modulators for massive WDM data transmission. Using 50-GHz on-chip EO combs pumped by two lasers, we achieve 46-channel (spanning 2.3 THz bandwidth) 45-GBaud dual-polarization (DP) 16-QAM transmission, delivering an aggregate data rate of 16.56 Tb/s (net 14.27 Tb/s) over a 98.84-km field-deployed link connecting two universities and two data centers in Hong Kong—validating its practical utility for inter-datacenter links. To illustrate the WDM grid flexibility of our scheme, we further interleave the comb into 100-GHz spacing via an on-chip de-multiplexer, demonstrating interleaved 10-channel, 80-GBaud WDM transmission at 6.4 Tb/s (net 5.5 Tb/s) over a 100-km fiber. Finally, to close the monolithic integration gap of existing EO comb demonstrations, we present a proof-of-concept on-chip WDM transmitter that integrates the comb generator with a high-speed modulator on a single chip, supporting 19-channel, 45-GBaud PAM-4 transmission at 1.71 Tb/s (net 1.47 Tb/s). Our demonstrations establish flat-top EO combs on TFLN as a powerful and scalable WDM solution for both coherent and IM/DD modalities, positioning this platform as a leading candidate for next-generation fully integrated optical transmitters.

## Results

### Conceptual framework of monolithically integrated flat-top EO comb-based WDM transmitters

Figure 1(a) illustrates an envisioned monolithically integrated EO comb-based WDM transmitter chip on TFLN for DCI applications. Flat-top EO comb is generated by cascading an amplitude modulator (AM) and a phase modulator (PM), eliminating the need for off-chip waveshapers. The comb's spectral coverage is governed by the number and wavelengths of the input carriers, as illustrated in the first inset, and can be flexibly tuned across O, C, and other telecommunication bands to accommodate diverse DCI applications. Its line spacing, defined by the RF source, can be precisely adjusted to match various commercial dense wavelength-division multiplexing (DWDM) grids (e.g., 50 or 100 GHz). The generated comb then passes through a de-multiplexing bank to interleave and separate it into distinct wavelength channels. In our proof-of-concept demonstration, we use an add-drop ring resonator to separate the even and odd channels, each featuring 100-GHz spacing, as illustrated in the second inset of Fig. 1(a). This method enables broadened signal bandwidths compatible with current 800G-per-wavelength DCI transmitters, and can be further scaled to support future 1.6T/3.2T-per-wavelength transmitters. In a full WDM transmitter system, individual comb lines can be further extracted by cascading multiple add-drop rings, or alternative approaches such as arrayed waveguide gratings (AWGs)[33–35], filter arrays based on Fabry-Perot cavities[46], or ring-assisted Mach-Zehnder interferometers (RAMZIs)[47]. The de-multiplexed comb lines then pass through monolithic integrated high-speed modulators for data encoding. Depending on the application scenario, in-phase/quadrature (IQ) modulators[37,38] or intensity modulators[25,26,36] can be used for coherent and IM/DD links, respectively. Finally, modulated signals can be recombined into a single fiber via an AWG or RAMZI array. This monolithic integration approach could potentially resolve the cross-platform challenges that limited prior comb-based systems, while also eliminating the need for numerous off-chip components. Figure 1(b) illustrates a representative 50-GHz EO comb spectrum featuring 46 channels across a 20 nm span with 6-dB flatness. Generated using two input carriers, the spectrum demonstrates robust performance for multi-carrier operation.

**Dual-carrier 50-GHz flat-top EO comb for inter-datacenter interconnects**

To establish the capability of our TFLN EO comb as a robust WDM source, we first conducted a comprehensive experimental investigation of a dual-carrier 50-GHz flat-top EO comb using the setup illustrated in Fig. 2(a). Two laser signals at 1548 nm and 1559.4 nm are combined and simultaneously injected into a TFLN EO comb chip with cascaded AM and PM [Fig. 2(c); see Supplementary Information Fig. S1 for details of the components]. The PM operates in a double-pass configuration[32] to double the modulation strength, which features a half-wave voltage ($V_\pi$) of approximately 4.5 V at 50 GHz. A 50-GHz RF signal is split, phase shifted, and fed to the two modulators, both of which employ capacitive-loaded electrodes [Fig. 2(d)] for low RF loss and impedance matching. At an amplified driving power of ~34 dBm on the PM (corresponding to a modulation depth of 3.5π), the resulting EO comb [Fig. 1(b)] achieves 6 dB flatness across 46 channels. Subsequently, the comb is modulated simultaneously using a commercial DP IQ modulator at 45-GBaud with QPSK and 16-QAM formats. Notably, the excellent spectral uniformity of the EO comb eliminates the need for a benchtop waveshaper or wavelength-selective switch (WSS) for power equalization. Figure 2(e) shows the modulated and amplified output spectra, where a slight degradation in power flatness (9.4 dB) is observed due to the non-uniform gain of the EDFA.

We then demonstrate ultrahigh-capacity WDM transmission at an aggregate rate of 16.56 Tb/s over a 98.84-km field-deployed link in Hong Kong. This fiber link connects two universities and two data centers, traversing diverse terrains including urban streets, mountains, and underwater tunnels across Victoria Harbour [Fig. 2(b)]. At the receiver, each channel is filtered, coherently detected, and processed offline. The received optical spectra [Fig. 2(f)] exhibit an additional 0.4 dB flatness degradation (left y-axis), primarily attributed to the two inline EDFAs compensating for >40 dB link loss. The power non-uniformity leads to a SNR variation of ~ 3 dB across channels [right *y*-axis, Fig. 2(f)]. Crucially, all 46 WDM channels maintain bit error ratios (BERs) below the soft-decision forward error correction (SD-FEC) threshold [Fig. 2(g)]. This yields a total data rate of 16.56 Tb/s for DP-16-QAM (net 14.27 Tb/s with 16% FEC overhead) and a near error-free rate of 8.28 Tb/s for DP-QPSK.

**On-chip interleaved 100-GHz flat-top EO comb for data-center interconnects**

To showcase the WDM grid flexibility and scalability of our scheme, we further combine the EO comb generator with an on-chip interleaver to produce a 100-GHz EO comb that facilitates 80-GBaud data transmission [Fig. 3(a)]. The optical interleaver is implemented using an add-drop microring resonator with a 100-GHz free spectral range (FSR). By aligning the comb lines with the ring resonances, the interleaver spatially separates the even and odd channels into the through and drop ports. Figure 3(b) shows the measured transmission spectra, confirming uniform filtering with extinction ratios exceeding 15 dB across a broad bandwidth. To ensure robustness against fabrication errors and thermal drift, we employ a strongly over-coupled pulley design, resulting in a relatively low loaded quality factor of $2.45 \times 10^4$ and a linewidth of 7.5 GHz [inset of Fig. 3(b)]. The resulting 100-GHz comb at the drop port achieves a 4.6 dB flatness across 10 channels, with a minimum side-mode suppression ratio of 15 dB. This interleaved comb is subsequently modulated with 80-GBaud QPSK and 16-QAM signals and transmitted over a 98.84-km field-deployed fiber link and a 100-km fiber spool, respectively, using a similar experimental setup as described earlier. Notably, our EO comb and on-chip interleaver demonstrate excellent wavelength agility, enabling operation at distinct wavelength bands without compromising performance. Using the same photonic chip, the DP 16-QAM test is performed at a center wavelength of 1534.3 nm [Fig.3(c)], whereas the DP QPSK signal is centered at 1556.65 nm (Supplementary Information Fig. S3). The received DP 16-QAM exhibit good power flatness, consistent SNR [Fig. 3(d)], and BERs below SD-FEC threshold [Fig. 3(e)], yielding a total data rate of 6.4 Tb/s (net 5.5 Tb/s). Figure 3(f) presents representative QPSK and 16-QAM constellations (see Supplementary Information Fig. S5 for full results). The signal performance under different baud rates is provided in Supplementary Information Fig. S4.

**Proof-of-concept on-chip WDM transmitter**

To demonstrate the feasibility of on-chip data transmission, we implement a proof-of-concept WDM transmitter by monolithically integrating the EO comb with a high-speed modulator. Figure 4(a) shows the measurement setup. The integrated modulator exhibits a 3-dB EO bandwidth exceeding 40 GHz, as shown by the EO $S_{21}$ response in Fig. 4(b). Its half-wave voltage ($V_\pi$) is measured to be 2.2 V using a 10 kHz triangular signal [Fig. 4(c)]. For the transmission experiment, we drive the comb generator's PM with a ~32 dBm on-chip RF signal at 50 GHz, producing a 19-channel comb with 6 dB flatness [Fig. 4(d), left]. This comb is subsequently modulated on-chip using a 45-GBaud PAM-4 signal, with a bias-tee employed to maintain the quadrature bias point. The modulated comb spectra are shown on the right of Fig. 4(d) (light purple lines), with the corresponding back-to-back received SNR plotted on the right y-axis (dark blue dots). Leveraging this monolithically integrated transmitter, we achieve a total data rate of 1.71 Tb/s (net 1.47 Tb/s). Figure 4(e) details the channel-wise BER results, while Fig. 4(f) presents representative eye diagrams (see Supplementary Fig. S7 for complete channel data).

**Discussion**

In summary, we present a scalable and versatile TFLN photonic architecture that monolithically integrates EO comb generation, on-chip (de-)multiplexing, and high-speed modulation for high-capacity WDM data transmission. Beyond demonstrating robust performance in both coherent and IM/DD formats, our work establishes flat-top EO combs as compact, efficient, and scalable building blocks for fully integrated WDM optical transmitters. The capability to support diverse modulation formats, flexible WDM grids, and seamless integration positions this platform as a compelling solution for next-generation chip-scale DCI systems.

While EO combs have demonstrated strong potential for high-capacity data transmission, a key barrier to widespread deployment remains their reliance on discrete external RF sources and high RF driving power to achieve broad comb spans. This challenge can be mitigated through co-packaged optics (CPO)[1], which allows for the integration of voltage-controlled oscillators (VCOs)[48], RF amplifiers[49], and our EO comb-based WDM transmitters on a single board. Alternatively, optoelectronic oscillators (OEO) based on TFLN modulators offer the prospect of EO comb generation without external RF sources [43,50]. To further reduce RF power requirements, our previously developed multi-loop phase modulators[27] and coplanar microwave resonator electrodes[31] can be adopted to achieve up to 16× power savings without compromising comb performance.

A second critical challenge is SNR, which is currently constrained by on-chip propagation and fiber-to-chip coupling losses. This limitation can be addressed through advanced design and fabrication techniques—for instance, using multi-layer inverse tapers to enhance mode overlap and minimize coupling loss[38]. Complementary to this, integrating on-chip TFLN waveguide amplifiers offers an effective approach to boost signal strength and enhance overall system performance[51,52]. Collectively, these future advances could be built upon the platform demonstrated in this work, enabling practical, high-performance, and fully integrated EO comb-based WDM transmitters for next-generation optical interconnects.

## Methods

### Device fabrication

Device fabrication begins with commercial 4-inch x-cut TFLN wafers (NANOLN), featuring a 500-nm LN film on a silicon substrate with a 4.7-μm buried oxide ($SiO_2$) layer. First, a $SiO_2$ hard mask is deposited via plasma-enhanced chemical vapor deposition (PECVD). Microstructures and optical waveguides are then patterned using an ASML UV stepper lithography system, which is capable of processing 1.5 cm×1.5 cm dies in sequence with a resolution of 500 nm. The patterns are transferred to the LN layer via inductively coupled plasma reactive-ion etching (ICP-RIE), followed by the deposition of a $SiO_2$ top cladding using PECVD. Finally, modulation electrodes, which consist of 500 nm of copper (Cu) capped with 50 nm gold (Au), are fabricated using photolithography, thermal evaporation, and lift-off processes.

**Details of comb generation**

16 dBm continuous wave (CW) light from an external cavity laser (ECL) is coupled into the x-cut LN chip via a polarization-maintained lensed fiber, with the fiber's slow axis aligned to the horizontal direction of the LN waveguide to excite the quasi-$TE_0$ mode. For dual-carrier comb generation, two wavelength-tunable ECLs are combined using a 3 dB polarization-maintained combiner prior to coupling. On-chip, the CW light is first modulated by the AM to carve a flat-top temporal pulse, which is then mapped to the frequency domain via a cascaded double-pass PM (through time-to-frequency conversion), yielding a flat-top EO comb[32]. The RF source dictates the comb line spacing, which is flexibly tunable. In this experiment, a fixed 50-GHz signal is provided by a phase-locked dielectric resonator oscillator (PLDRO) and split into two channels via a 3-dB RF splitter. Both signals are applied to the chip using RF probes (GGB Industries Inc.). An RF phase shifter (Sage Millimeter, STP-18-19-M1) compensates for phase path differences between the modulators. In the AM path, a bias-tee (Marki Microwave, BT-0050) sets the bias point for flat-top pulse generation, while an amplifier (SHF S807) ensures sufficient modulation depth. In the PM path, the signal is amplified to ~34 dBm (Talent Microwave, TMPA-040060-3036-19), with a tunable attenuator (Sage Millimeter, STA-30-19-M2A-WPC) allowing flexible control over the comb span (number of lines).

**Details of the coherent experiment**

The comb tones are first simultaneously amplified by a polarization-maintained EDFA to 13 dBm, and directed to a bulk lithium niobate DP IQ modulator (Fujitsu FTM7992HM, 35 GHz@3 dB). The modulator is stabilized at the null point using an automatic bias controller (Plugtech MBC-DPIQ-02). All comb tones are simultaneously modulated with DP-16QAM or DP-QPSK signals generated by an arbitrary waveform generator (Keysight M8194A, 120 GSa/s, 45 GHz bandwidth), utilizing a raised-cosine pulse shape with a roll-off factor of 0.05. The RF drive signals are boosted by electrical amplifiers (SHF 807C, 55 GHz bandwidth) prior to modulation.

Following optical amplification, the modulated WDM signal is launched into the transmission link. For the QPSK test, we utilize a 98.84-km field-deployed fiber network (HKCOLO), while the 16-QAM test employs a 100-km standard G.652 fiber spool (0.2 dB/km). The field link [Fig. 2(b)] originates at The Hong Kong Polytechnic University (PolyU), traveling 26.9 km to the Cloud

Computing Complex (CCC) in Tseung Kwan O—a Tier-4 data center serving the Asia-Pacific financial corridor. It then loops back 23.2 km to the City University of Hong Kong (CityU), where it is amplified by an inline EDFA. From CityU, the signal crosses Victoria Harbour, traversing 28.74 km to the Sino Favour Centre (SFC) in Chai Wan, a major carrier colocation hub, before completing a 20-km return path to PolyU for detection.

At the receiving side, the WDM signal is pre-amplified and filtered by a tunable filter (Coherent, Waveshaper, 4000A) and mixed with a local oscillator (16 dBm, served by a wavelength-tunable ECL) in a DP 90° optical hybrid. The mixed signal is coherently detected by four balanced photodiodes (BPDs), down-converted to the baseband, and then sampled by a digital oscilloscope (Keysight UXR804A, 256 GSa/s, 80 GHz@3 dB) for offline processing. The sampled signal passes through a series of digital signal processing (DSP) stages consisting of chromatic dispersion (CD) compensation, polarization demultiplexing using constant modulus algorithm (CMA) and radius directed equalization (RDE) by 2x2 complex-valued Multi-Input-Multi-Output (MIMO) equalizer, $4^{th}$-power carrier frequency recovery (CFR), carrier phase recovery (CPR) by blind phase search[53] followed by symbol decision and bit error counting. The demodulation process is conducted throughout each WDM channel by adjusting the waveshaper's center wavelength and the local oscillator wavelength to that of the channel under test (CUT).

**Details of the IMDD experiment**

The IMDD experiment is conducted on a back-to-back (B2B) configuration. The 45-GBaud PAM-4 signal is generated by Keysight 8194A and amplified by an electrical amplifier, before being loaded onto the on-chip intensity modulator through an RF probe, whose bias voltage is manually controlled using a bias-tee. A raised-cosine waveform with a roll-off factor of 0.01 is used as the basic waveform for modulation. The modulated WDM signal is amplified by an EDFA and sent to the waveshaper, which filters out the CUT and avoids neighbor channel beating at detection. The CUT is detected by a transimpedance amplifier (TIA)-equipped PD (Finisar, XPRV2022A, 33 GHz@3 dB). The detected signal is sampled by Keysight UXR804A and processed offline. An absolute-term-based nonlinear feed-forward equalization algorithm[54] is used to equalize inter-symbol interference (ISI) in the received signal. The demodulation process is conducted throughout each WDM channel by adjusting the waveshaper's center wavelength to that of the CUT.

**Funding.** National Natural Science Foundation of China (62525507); Research Grants Council, University Grants Committee (CityU11212721, CityU 11213125, PolyU15225423, JRFS2526-1S01, C5078-24G); Croucher Foundation (9509005); City University of Hong Kong (9610682).
**Acknowledgments.** We are grateful for the generous donation of the deployed fiber link by HKCOLO/Telehouse.
**Conflict of interest.** K.Z. is the founder and shareholder of RhinoptiX Technology Ltd., a company developing thin-film lithium niobate technologies. The remaining authors declare no competing interests.
**Data availability.** Data underlying the results presented in this paper are not publicly available at this time but may be obtained from the authors upon reasonable request.

**References**

1. Tan, M. *et al.* Co-packaged optics (CPO): status, challenges, and solutions. *Front. Optoelectron.* **16**, 1 (2023).
2. Cheng, Q., Bahadori, M., Glick, M., Rumley, S. & Bergman, K. Recent advances in optical technologies for data centers: a review. *Optica* **5**, 1354 (2018).
3. Xie, C. & Zhang, B. Scaling Optical Interconnects for Hyperscale Data Center Networks. *Proc. IEEE* **110**, 1699–1713 (2022).
4. Zhou, X., Urata, R. & Liu, H. Beyond 1 Tb/s Intra-Data Center Interconnect Technology: IM-DD OR Coherent? *J. Light. Technol.* **38**, 475–484 (2020).
5. Geng, Y. *et al.* Coherent optical communications using coherence-cloned Kerr soliton microcombs. *Nat. Commun.* **13**, 1070 (2022).
6. Pfeifle, J. *et al.* Coherent terabit communications with microresonator Kerr frequency combs. *Nat. Photonics* **8**, 375–380 (2014).
7. Wang, W. *et al.* Free-space terabit/s coherent optical links via platicon frequency microcombs. *eLight* **5**, 8 (2025).
8. Zhang, X. *et al.* High-coherence parallelization in integrated photonics. *Nat. Commun.* **15**, 7892 (2024).
9. Fülöp, A. *et al.* High-order coherent communications using mode-locked dark-pulse Kerr combs from microresonators. *Nat. Commun.* **9**, 1598 (2018).
10. Pirmoradi, A. *et al.* Integrated multi-port multi-wavelength coherent optical source for beyond Tb/s optical links. *Nat. Commun.* **16**, 6387 (2025).
11. Rizzo, A. *et al.* Massively scalable Kerr comb-driven silicon photonic link. *Nat.*

*Photonics* **17**, 781–790 (2023).

12. Shu, H. *et al.* Microcomb-driven silicon photonic systems. *Nature* **605**, 457–463 (2022).
13. Marin-Palomo, P. *et al.* Microresonator-based solitons for massively parallel coherent optical communications. *Nature* **546**, 274–279 (2017).
14. Yang, K. Y. *et al.* Multi-dimensional data transmission using inverse-designed silicon photonics and microcombs. *Nat. Commun.* **13**, 7862 (2022).
15. Liu, Y. *et al.* Parallel wavelength-division-multiplexed signal transmission and dispersion compensation enabled by soliton microcombs and microrings. *Nat. Commun.* **15**, 3645 (2024).
16. Jørgensen, A. A. *et al.* Petabit-per-second data transmission using a chip-scale microcomb ring resonator source. *Nat. Photonics* **16**, 798–802 (2022).
17. Shao, W. *et al.* Terabit FSO communication based on a soliton microcomb. *Photonics Res.* **10**, 2802 (2022).
18. Raja, A. S. *et al.* Ultrafast optical circuit switching for data centers using integrated soliton microcombs. *Nat. Commun.* **12**, 5867 (2021).
19. Kemal, J. N. *et al.* 32QAM WDM transmission at 12 Tbit/s using a quantum-dash mode-locked laser diode (QD-MLLD) with external-cavity feedback. *Opt. Express* **28**, 23594 (2020).
20. Lu, Z. *et al.* InAs/InP Quantum Dash Semiconductor Coherent Comb Lasers and their Applications in Optical Networks. *J. Light. Technol.* **39**, 3751–3760 (2021).
21. Liu, G. *et al.* InAs/InP quantum dot mode-locked laser with an aggregate 12.544

Tbit/s transmission capacity. *Opt. Express* **30**, 3205 (2022).

22. Mao, Y. *et al.* Ultralow Noise and Timing Jitter Semiconductor Quantum-Dot Passively Mode-Locked Laser for Terabit/s Optical Networks. *Photonics* **9**, 695 (2022).
23. Bernal, S. *et al.* 12.1 terabit/second data center interconnects using O-band coherent transmission with QD-MLL frequency combs. *Nat. Commun.* **15**, 7741 (2024).
24. Fujiwara, M. *et al.* Optical carrier supply module using flattened optical multicarrier generation based on sinusoidal amplitude and phase hybrid modulation. *J. Light. Technol.* **21**, 2705–2714 (2003).
25. Wang, C. *et al.* Integrated lithium niobate electro-optic modulators operating at CMOS-compatible voltages. *Nature* **562**, 101–104 (2018).
26. Zhang, Y. *et al.* Systematic investigation of millimeter-wave optic modulation performance in thin-film lithium niobate. *Photonics Res.* **10**, 2380 (2022).
27. Zhang, K. *et al.* A power-efficient integrated lithium niobate electro-optic comb generator. *Commun. Phys.* **6**, 17 (2023).
28. Zhang, M. *et al.* Broadband electro-optic frequency comb generation in a lithium niobate microring resonator. *Nature* **568**, 373–377 (2019).
29. Wen, H., Wang, J., Zhu, L., Shi, L. & Zhang, X. Broadband and flat-top integrated electro-optic frequency combs on a thin-film lithium niobate platform. *Opt. Lett.* **50**, 3692 (2025).
30. Wang, J. *et al.* Highly tunable flat-top thin-film lithium niobate electro-optic frequency comb generator with 148 comb lines. *Opt. Express* **33**, 23431 (2025).

31. Chen, Z. *et al.* Microwave-resonator-enabled broadband on-chip electro-optic frequency comb generation. *Photonics Res.* **13**, 426 (2025).

32. Yu, M. *et al.* Integrated femtosecond pulse generator on thin-film lithium niobate. *Nature* **612**, 252–258 (2022).

33. Yi, J. *et al.* Anisotropy-free arrayed waveguide gratings on X-cut thin film lithium niobate platform of in-plane anisotropy. *Light Sci. Appl.* **13**, 147 (2024).

34. Liu, H. *et al.* First demonstration of lithium niobate photonic chip for dense wavelength-division multiplexing transmitters. *Adv. Photonics* **6**, 066001–066001 (2024).

35. Wang, Z. *et al.* On-Chip Arrayed Waveguide Grating Fabricated on Thin-Film Lithium Niobate. *Adv. Photonics Res.* **5**, 2300228 (2024).

36. Chen, Y. *et al.* Mono-drive single-sideband modulation via optical delay lines on thin-film lithium niobate. *Optica* **12**, 666 (2025).

37. Xu, M. *et al.* High-performance coherent optical modulators based on thin-film lithium niobate platform. *Nat. Commun.* **11**, 3911 (2020).

38. Xu, M. *et al.* Dual-polarization thin-film lithium niobate in-phase quadrature modulators for terabit-per-second transmission. *Optica* **9**, 61 (2022).

39. Zhu, S. *et al.* Integrated lithium niobate photonic millimetre-wave radar. *Nat. Photonics* **19**, 204–211 (2025).

40. Feng, H. *et al.* Integrated lithium niobate microwave photonic processing engine. *Nature* **627**, 80–87 (2024).

41. Guo, Q. *et al.* Ultrafast mode-locked laser in nanophotonic lithium niobate. *Science*

**382**, 708–713 (2023).

42. Yu, M. *et al.* Integrated electro-optic isolator on thin-film lithium niobate. *Nat. Photonics* **17**, 666–671 (2023).
43. Tao, Z. *et al.* Ultrabroadband on-chip photonics for full-spectrum wireless communications. *Nature* **645**, 80–87 (2025).
44. Zhang, C. *et al.* High-fidelity sub-petabit-per-second self-homodyne fronthaul using broadband electro-optic combs. *Nat. Commun.* **15**, 6621 (2024).
45. Liu, X. *et al.* Thin-film lithium niobate-based electro-optic comb cloning for self-homodyne coherent communication. *Opt. Lett.* **49**, 3238 (2024).
46. Zhu, M. *et al.* Multichannel Lithium-Niobate-On-Insulator Photonic Filter for Dense Wavelength-Division Multiplexing. *ACS Photonics* **12**, 588–596 (2025).
47. Feng, H. *et al.* On-chip optical vector analysis based on thin-film lithium niobate single-sideband modulators. *Adv. Photonics* **6**, 066006–066006 (2024).
48. Issakov, V. The State of the Art in CMOS VCOs: Mm-Wave VCOs in Advanced CMOS Technology Nodes. *IEEE Microw. Mag.* **20**, 59–71 (2019).
49. Hamid, S. S. *et al.* A State-of-the-Art Review on CMOS Radio Frequency Power Amplifiers for Wireless Communication Systems. *Micromachines* **14**, 1551 (2023).
50. Ma, R. *et al.* Ka-band thin film lithium niobate photonic integrated optoelectronic oscillator. *Photonics Res.* **12**, 1283 (2024).
51. Chen, Z. *et al.* Efficient erbium-doped thin-film lithium niobate waveguide amplifiers. *Opt. Lett.* **46**, 1161 (2021).
52. Zhou, J. *et al.* On-Chip Integrated Waveguide Amplifiers on Erbium-Doped Thin-

Film Lithium Niobate on Insulator. *Laser Photonics Rev.* **15**, 2100030 (2021).

53. Pfau, T., Hoffmann, S. & Noe, R. Hardware-Efficient Coherent Digital Receiver Concept With Feedforward Carrier Recovery for M-QAM Constellations. *J. Light. Technol.* **27**, 989–999 (2009).

54. Wu, X., Zhang, J., Lau, A. P. T. & Lu, C. Low-complexity absolute-term based nonlinear equalizer with weight sharing for C-band 85-GBaud OOK transmission over a 100-km SSMF. *Opt. Lett.* **47**, 1565 (2022).

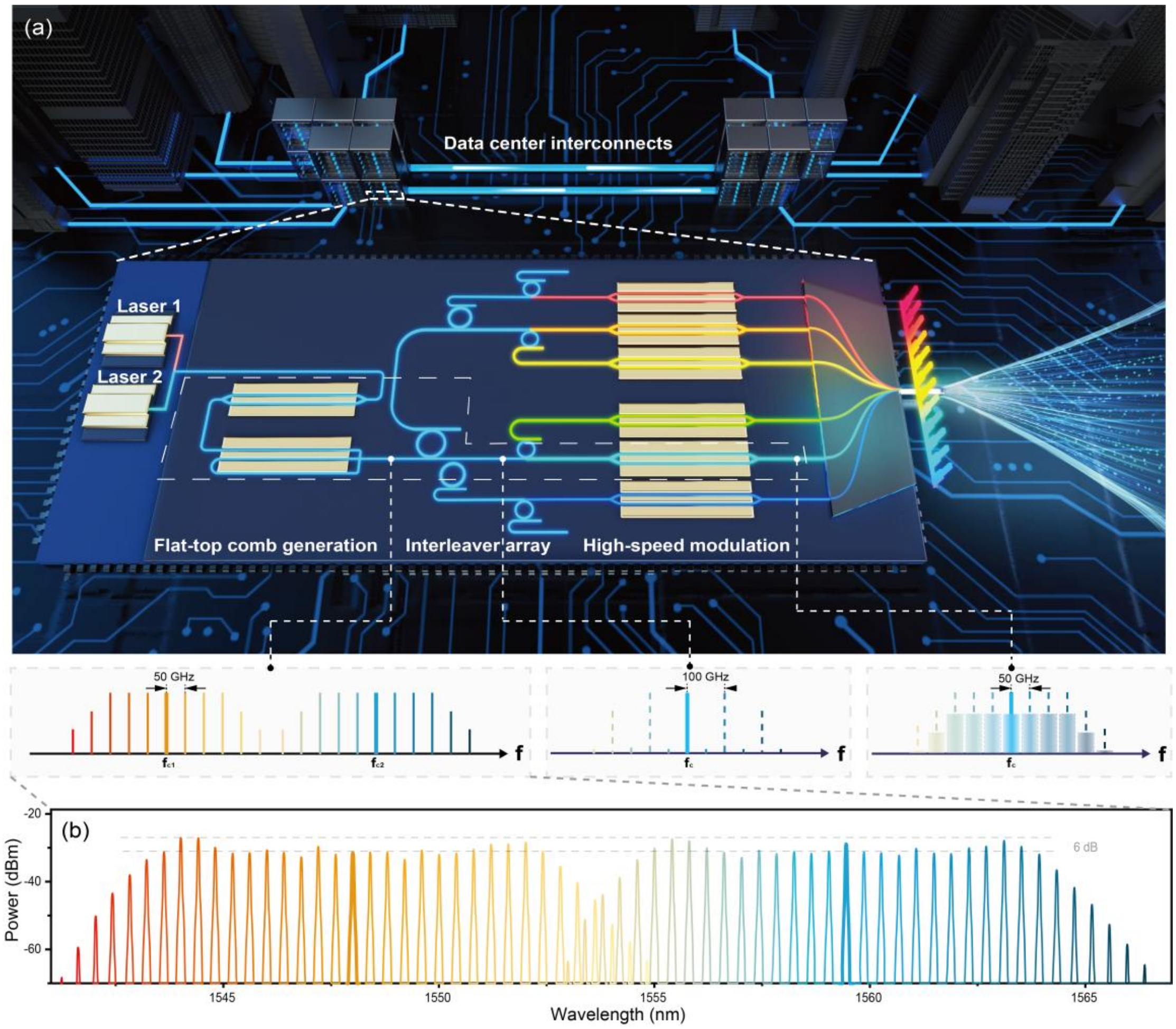

**Fig. 1.** (a) Conceptual schematic of a monolithically integrated flat-top electro-optic (EO) comb-based wavelength-division multiplexing (WDM) transmitter on thin-film lithium niobate (TFLN) platform for data center interconnects (DCI). Insets correspond to optical spectra at various locations of the proposed architecture. (b) Measured spectrum of a dual-carrier 50-GHz EO comb comprising 46 flat-top channels.

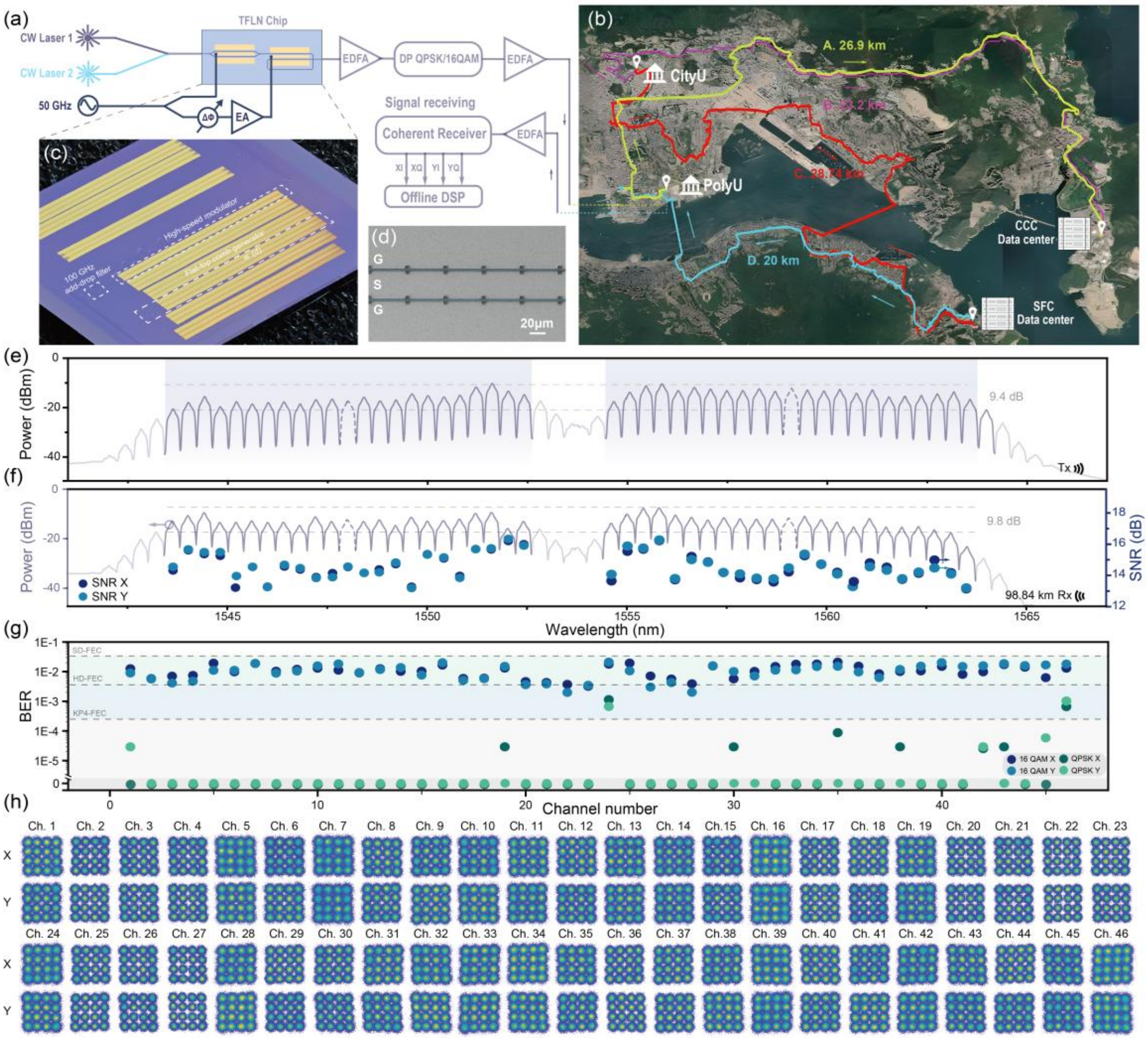

**Fig. 2.** (a) Experimental setup for dual-carrier flat-top EO comb-based WDM data transmission. (b) Map illustration of the 98.84-km field-deployed fiber link connecting two universities and two data centers in Hong Kong. (c) Microscope image of the fabricated TFLN chip. (d) Scanning electron microscope image of the capacitive-loaded electrodes. (e) Measured optical spectrum of the dual-carrier EO comb carrying 45-GBaud modulation signals. (f) Received spectrum (left *y*-axis) and corresponding channel-wise signal-to-noise ratios (SNRs, right *y*-axis) after transmission over the deployed link. (g) Measured bit error ratios (BERs) for DP 16QAM (dark/light blue: X/Y polarization) and DP QPSK (dark/light green: X/Y polarization). (h) Received constellations for all 46 DP 16QAM channels.

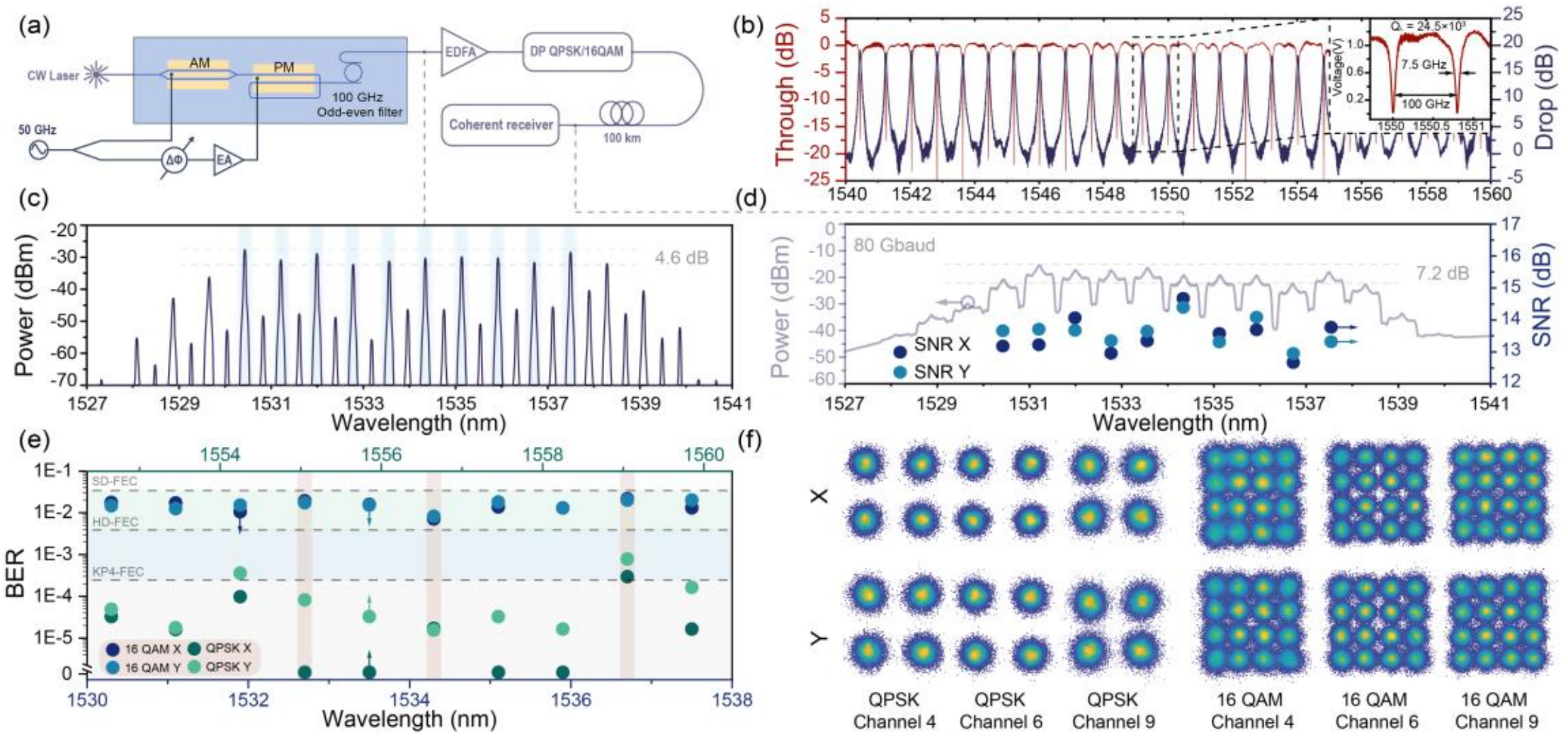

**Fig. 3.** (a) Schematic of the data transmission measurement setup using an interleaved 100-GHz flat-top EO comb. (b) Transmission spectra of the 100-GHz add-drop interleaver at the through port (red) and drop port (blue). Inset provides a zoom-in view of the through-port spectrum around 1550 nm. (c) Output spectra of the interleaved 100-GHz comb at the drop port. (d) Received optical spectra (left *y*-axis) and corresponding channel-wise SNRs (right *y*-axis) of DP 16QAM after transmission over 100 km fiber. (e) Tested BERs for DP 16QAM (dark/light blue: X/Y polarization) and DP QPSK (dark/light green: X/Y polarization). (f) Received constellations in representative channels.

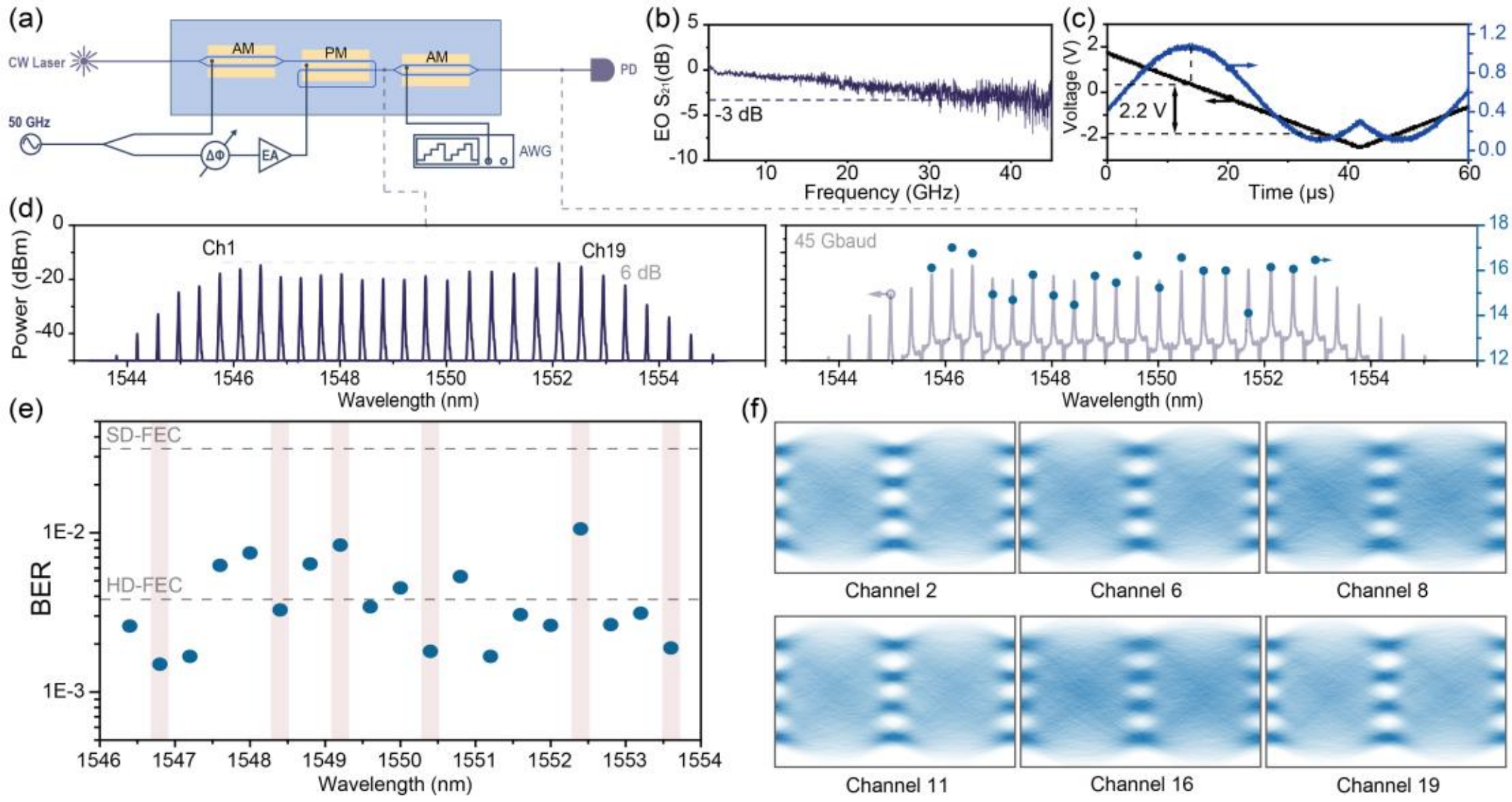

**Fig. 4.** (a) Schematic of the data transmission setup using a monolithically integrated flat-top EO comb and on-chip intensity modulator. (b-c) Measured (b) EO $S_{21}$ and (c) half-wave voltage ($V_\pi$) of the monolithically integrated intensity modulator. (d) Optical spectra of the generated 50-GHz flat-top EO comb (left) and the modulated comb after on-chip PAM-4 modulation (right). The corresponding back-to-back received SNR for each channel is plotted on the right Y-axis (dark blue dots). (e) BER results for all channels. (f) Eye diagrams of representative channels.